# Heterostructure Design in Two-Dimensional Perovskites by Sequential Recrystallization

*Mehrdad Faraji[1,2], Alexander Schleusener[1], Sirous Khabbaz Abkenar[3], Andrea Griesi[3], Mattia Lizzano[3], Sudhir Kumar Saini[1], Aswin V. Asaithambi[4], Liberato Manna[4], Matteo Lorenzoni[5], Mirko Prato[5], Giorgio Divitini[3], and Roman Krahne[1]*

*[1]Optoelectronics, Italian Institute of Technology (IIT), Via Morego 30, 16163 Genoa, Italy*
*[2]Dipartimento di Chimica e Chimica Industriale, Università degli Studi di Genova, Via Dodecaneso, 31, Genova 16146, Italy*
*[3]Electron Spectroscopy and Nanoscopy, Italian Institute of Technology (IIT), Via Morego 30, 16163 Genoa, Italy*
*[4]Nanochemistry, Italian Institute of Technology (IIT), Via Morego 30, 16163 Genoa, Italy*
*[5]Material Characterization Facility, Italian Institute of Technology (IIT), Via Morego 30, 16163 Genoa, Italy*

*Corresponding authors email: alexander.schleusener@iit.it, roman.krahne@iit.it*

**Abstract**

Low-dimensional metal halide perovskites provide exciting opportunities to fabricate new semiconductor materials. Semiconductor technology relies on electronic heterojunctions, and cost-efficient and flexible approaches to realize functional heterostructures are of fundamental importance. Lateral heterostructures define the energy landscape in the plane of the semiconducting lattice in such 2D materials, representing an ideal platform to tailor energy barriers and to control charge carrier flow. Here, we demonstrate a versatile one-pot synthesis to fabricate a large variety of 2D perovskite heterostructures based on different halides and/or metal cations. Exploiting sequential crystallization of different 2D perovskites, and playing with the composition and injection events of the materials, enables the design of diverse heterostructure architectures including multiple heterojunctions. We obtain crystalline quality of the heterojunctions, multicolor emission, and optical coupling between the different heterostructure regions. We foresee that the design freedom of our method will stimulate the development of novel optoelectronic devices where electronic band engineering is crucial.

## 1. Introduction

Two-dimensional layered metal-halide perovskites (2DLPs) are a rapidly developing material family in which semiconducting metal-halide octahedral layers are separated by bulky organic cations.[1–5] They combine strong excitonic effects and high refractive indices with structural and compositional tunability, making them attractive for optoelectronic and photonic applications.[6–12] A key challenge for such applications is the rational control of local composition and band gap, which governs energy and charge carrier flow in devices.[13–16] Semiconductor heterostructures are the pivotal element in optoelectronics, and pioneering works have reported the implementation of lateral and vertical heterojunctions in 2DLPs.[17,18] Vertical stacking or self-assembly of 2D materials offers one route to engineer the band alignment, but it primarily modulates the out-of-plane coupling.[19,14,20,21] However, in 2DLPs, the typically insulating organic spacer cations largely decouple adjacent layers, limiting interlayer interaction.[22,23] Similar to transition-metal dichalcogenides, lateral heterostructures in which composition varies in the in-plane direction offer a promising route toward spatially defined junctions, energy funnels, and multifunctional interfaces integrated within a single crystal.[24–27] Achieving high junction quality, ideally through epitaxial interfaces, is therefore critical. The soft lattice of 2DLPs can sustain such interfaces even in the presence of considerable lattice mismatch, particularly along the out-of-plane direction, where organic spacers decouple the inorganic slabs.[28,29]

Post-synthetic methods such as solution-based halide exchange have proven effective for creating macroscopically homogeneous lateral junctions, but are limited to halide substitution because the metal cations are deeply embedded in the structure.[30–33] On the microscopic scale, these junctions often suffer from poor crystallinity at the interface and halide-dependent limitations, resulting in a small number of potential heterostructures.[31] Epitaxial growth of 2DLP heterostructures has been demonstrated in substrate-constrained, higher-temperature, multi-step processes that require careful solvent and temperature control, but such methods lack versatility and are challenging to generalize.[29,34–36]

We introduce a one-pot, solution-based crystallization strategy that enables the rational design of lateral heterostructures in 2DLPs, offering flexibility in both material composition and heterostructure architecture. Our method exploits the pronounced differences in solubility of 2DLPs in polar solvents such as acetonitrile (ACN) and γ-butyrolactone (GBL). Rather than relying solely on solvent evaporation to induce crystallization[34,36], our approach provides additional control over the crystallization process by deliberately and sequentially injecting an antisolvent, such as toluene or dichlorobenzene (DCB). This controlled manipulation of the solution composition and supersaturation enables the sequential crystallization of distinct 2DLP compositions within a single reaction vessel, thereby providing a versatile one-pot route toward lateral heterostructures.

This method enables several heterostructure fabrication strategies. By dissolving two 2DLPs with different halide compositions in the starting solution of the reaction, we obtain core–frame structures consisting of alloyed phases with graded junctions. By instead dissolving only one material at the beginning and injecting the second 2DLP material solution at a later stage of antisolvent addition (sequential injection), we achieve halide heterostructures with phase-pure cores and nearly phase-pure frames. Here, the frame composition can be further tuned by adjusting the halide ratio of the injected solution, thereby enabling in-plane band gap engineering. In addition to conventional core–frame architectures, we observe the formation of distinct triptych-shaped structures (a central rectangle with two rectangular side panels), where the secondary phase grows preferentially along the long edges of rectangular core crystals. We demonstrate optical coupling across the heterojunction interface by selectively exciting the bromide phase core and detecting the emission from the iodide phase frame. Finally, we extend our method to more complex systems, including a novel triple-halide (Cl–Br–I) heterostructures within a single microcrystal, and to core-shell structures with Pb as metal cation in the core, and Ag and Bi in the shell.[35] Our findings establish a generalizable framework for the solution-based synthesis of lateral heterostructures in 2DLPs. Our approach opens new pathways for designing band gap-engineered optoelectronic and photonic devices that enable charge transfer and energy flow management, and which have the potential to drive innovations in photocatalysis, multicolor light emission, and photosensing.[37–39] Furthermore, such heterojunction microcrystals can be an ideal platform for fundamental studies of charge transfer and exciton dynamics in hybrid semiconductors.[40–42]

## 2. Results and Discussion

Figure 1a illustrates the concept of our one-pot sequential growth strategy: pre-synthesized two-dimensional layered perovskite powders[43] (SI Section 1, **Figs. S1-S2**) are dissolved in a solvent like ACN (in the following called precursor solution, see Table S2), and recrystallization is triggered by the addition of an antisolvent. This recrystallization from 2DLP powders offers the advantage of starting from phase-pure materials while avoiding the formation of unwanted side products (such as organic residues from precursor salts) during crystallization. We recently employed a similar strategy for the direct growth of well-defined 2DLP microcrystals that act as open optical cavities owing to their nearly atomically flat surfaces.[11] Tuning the antisolvent/solvent ratio over time by the controlled addition of the antisolvent enables control over the crystallization kinetics **(Fig. S3)**. To obtain heterostructures, only one material can be dissolved, and a second material can be either added after a certain period of time (sequential injection), or both materials can be dissolved in the mixture from the beginning. In the latter case, the material with the lowest solubility will crystallize first, and the delayed crystallization of the material with higher solubility results in core-frame or triptych-shaped architectures of heterostructured microcrystals. In this process, we effectively suppress homogeneous nucleation by slow antisolvent injection that increases the saturation of the solution gradually. The experimentally determined different solubility of 2DLPs with different halides and metal cations is reported in Figure 1b. With the sequential injection strategy, we can also grow more complex systems, such as core/frame/frame structures with three different halides. To demonstrate the large versatility of this method we show scanning transmission electron microscopy (STEM) and energy dispersive X-ray spectroscopy (EDX) images of microcrystals with different heterostructure compositions in Figures 1c-g: (c) core-frame chloride-bromide ($PEA_2PbCl_4$-$PEA_4PbBr_4$), (d) triptych-shaped bromide-iodide ($PEA_2PbBr_4$-$PEA_2PbI_4$), (e) triple halide chloride-bromide-iodide core/frame/triptych ($PEA_2PbCl_4$-$PEA_2PbBr_4$-$PEA_2PbI_4$), (f) alloyed $Br_x/I_y$ and $Br_{x'}/I_{y'}$ (with x>x') phases with different composition in core/frame and dumbbell architectures, and (g) heterostructures with $PEA_2PbBr_4$ in the center and the double perovskite $PEA_4AgBiBr_8$ at the edges.[35] To gain deeper insights into the growth process, we will focus first on the fabrication of heterostructures starting from a mixture of $PEA_2PbBr_4$ and $PEA_2PbI_4$ microcrystals dissolved in ACN (Figure 2a).

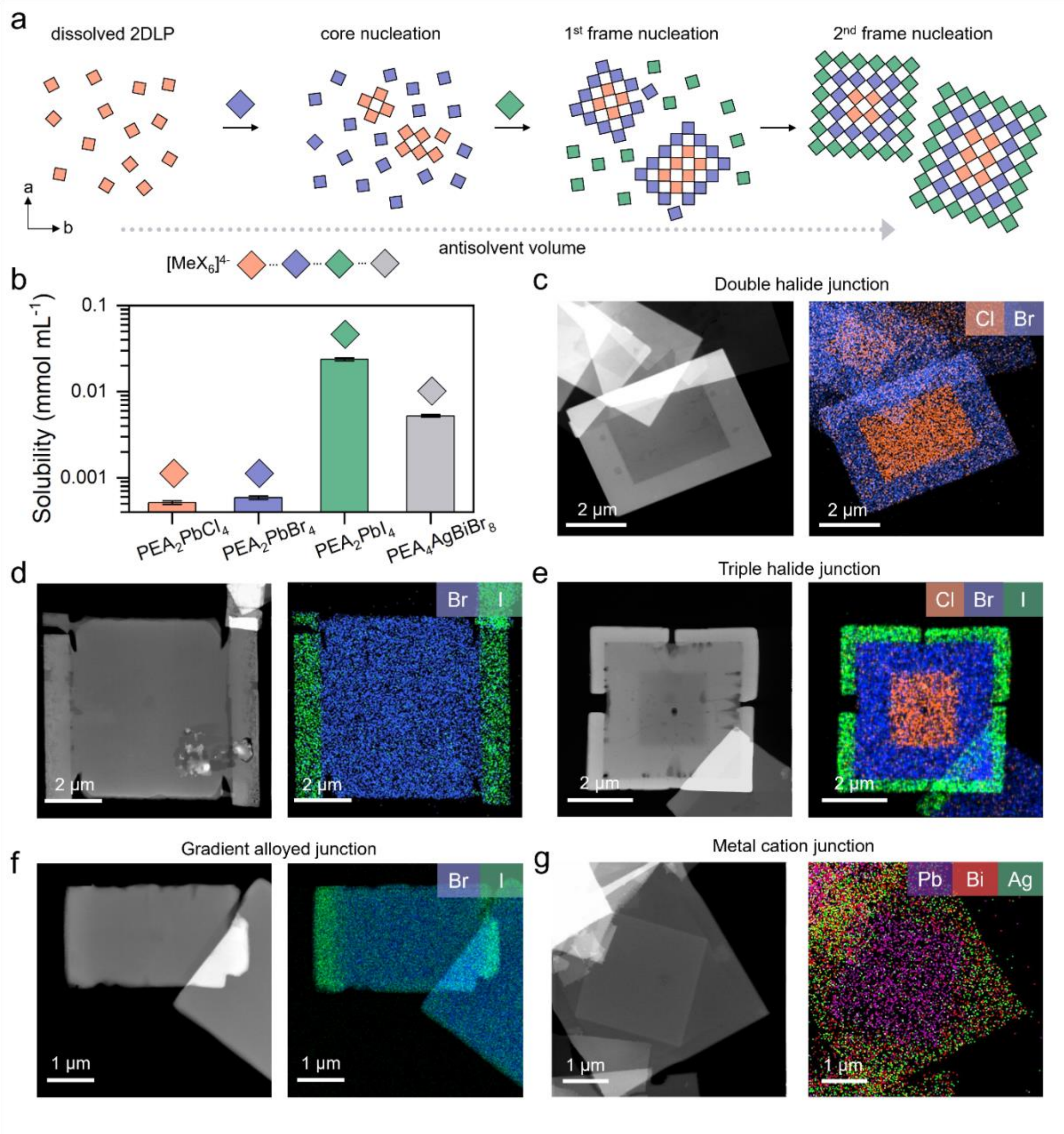

**Figure 1.** a) Illustration of the sequential crystallization process and how it leads to heterostructured microcrystals with core-frame (or triptych-shaped) architecture. b) Solubility of 2D perovskite microcrystals in the ACN as solvent determined by ICP. c-g) STEM and EDX images showing heterostructured microcrystals with different compositions: c) core-frame chloride-bromide ($PEA_2PbCl_4$-$PEA_2PbBr_4$), d) triptych-shaped bromide-iodide ($PEA_2PbBr_4$-$PEA_2PbI_4$), e) triple halide chloride-bromide-iodide core/frame/triptych ($PEA_2PbCl_4$-$PEA_2PbBr_4$-$PEA_2PbI_4$), f) alloyed $Br_x/I_y$ and $Br_{x'}/I_{y'}$, g) $PEA_2PbBr_4$-$PEA_4AgBiBr_8$.

### 2.1 Heterojunctions of alloyed composition by growth from mixtures

When initiating the reaction from mixtures of dissolved bromide and iodide 2DLP microcrystals, we find that the obtained heterostructures are composed of alloyed phases, that is, of a bromide-rich core and an iodide-rich phase at the edges. The compositions can be tuned by the relative concentrations of the different materials in the starting mixture (**Table S3**), which provides an opportunity to tailor the band gap and the heterojunction profile within the limits set by the phase-pure materials. The PL and absorption spectra in Figure 2b demonstrate such control on the absorption edge and PL center wavelength. We note that the emergence of new bands cannot be explained by a simple superposition of the optical responses of the two pristine materials, and the heterostructure formation is therefore attributed to alloyed phases of distinct Pb-Br-I solvent complexes in solution, which likely predetermine the core–frame composition of the resulting heterostructures (**Fig. S4**).

Figure 2c shows a series of hyperspectral confocal PL images that evidence the core-frame heterostructure architecture and the change in emission color for different I:Br material ratios in the starting mixture (see **Fig. S5** for lower magnification confocal imaging and SEM images, and **Figs. S6-S8** for additional XRD, hyperspectral confocal, and STEM-EDX data). In-situ monitoring of the PL during the crystallization process (Figure 2d-e) provides direct information on the microcrystal band gap (and therefore halide composition) via the spectral position of the PL maximum,[31] as well as on the different crystallization events through the evolution of the integrated PL intensity. Here the onset of the PL signal at 2.5 mL of antisolvent marks the start of the crystallization of a bromide-rich phase (PL centered around 485 nm) that remains stable up to ca. 9 mL (stage 1). The pronounced shift of the PL maximum from 485 nm to 500 nm and the increase of the PL signal in stage 2 indicates the progressive formation of halide alloys with differing compositions, resulting in a graded junction region as confirmed by STEM-EDX (Figure 2f-g). In stage 3 (12.5-18 mL), the PL position stabilizes at about 500 nm, while the PL intensity increases, which we attribute to the frame growth of an iodide-rich phase with stable composition. This interpretation was also confirmed by stopping the reactions at certain volumes of injected antisolvent volume (3.3, 6.7, 10.0 and 13.4 mL) to determine the size and spatial distribution of the blue and green PL signal (See **Fig. S9**). The increase in PL intensity (with wavelength corresponding to the frame material) reflects efficient energy transfer from the core to the frame region (Figure 2e). The local material composition across the heterojunction can be obtained using STEM-EDX, as shown in Figure 2f, which reveals a gradual decrease in the bromine signal towards the edge, correlating with an increase in the iodine signal, while the Pb signal remains constant. We conclude that the junction is characterized by a gradual change in composition that extends over several hundred nanometers. The homogeneous morphology of the heterojunction microcrystals points to an epitaxial (single-crystalline) interface. Such a graded shell architecture effectively compensates for lattice strain and typically results in a low density of defects.[44]

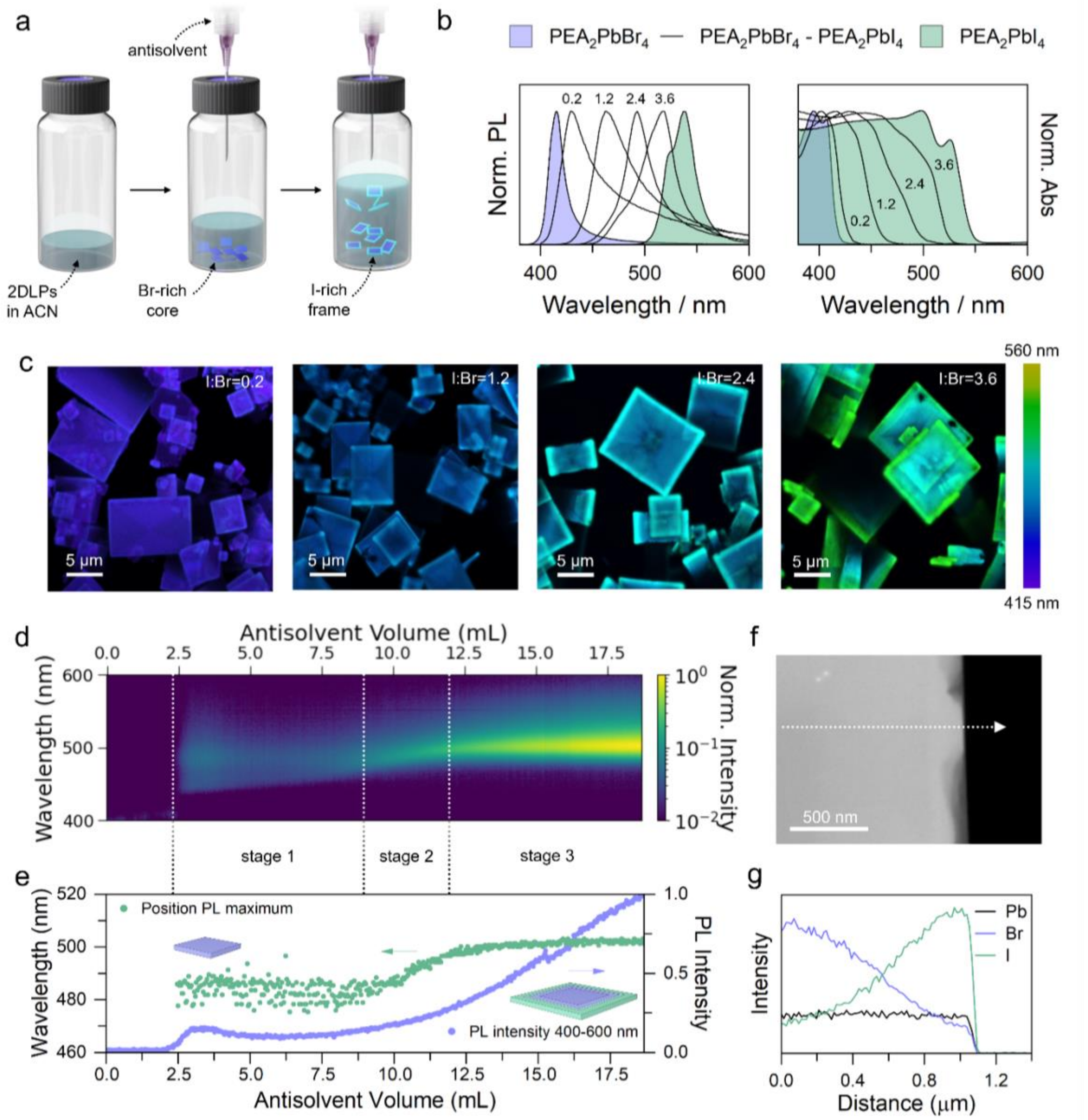

**Figure 2. Heterojunctions in 2D layered perovskites from mixed-halide precursor solutions**. a) Illustration of the growth process. b) PL ($\lambda_{ex}$=350 nm) and absorption spectra recorded from microcrystals fabricated with different I:Br ratios, demonstrating that the optical band gap of the mixed phases can be tuned by the molar composition of the dissolved materials. c) Confocal hyperspectral PL images of heterostructure microcrystals composed of alloyed $PEA_2PbBr_x/I_y$ phases with different composition ($\lambda_{ex}$=400 nm). d) Tracking the crystallization kinetics by in situ PL monitoring ($\lambda_{ex}$=365 nm) during the growth process from mixed precursor solution (1:1.2) with toluene as antisolvent (1 mL $min^{-1}$). e) Extracted position of the PL

maximum and integrated PL intensity (400-600 nm). f) High-Angle Annular Dark Field (HAADF)-STEM image of the heterojunction (I:Br=2.4) at the junction of a core-frame microcrystal, and g) EDX line scans for Pb, Br and I intensity along the indicated arrow in f).

### 2.2 Sequential injection: nearly phase-pure heterostructures with sharp interfaces and triple-halide architectures

Instead of starting the growth process from mixtures of different 2D perovskite materials, one can also dissolve a single material in a solvent like ACN or γ-butyrolactone (GBL), and inject another (dissolved) 2DLP material at a later stage, i.e., at higher antisolvent/solvent ratios. This method (depicted in Figure 3a) has two advantages: the initially grown crystals are phase-pure (as no other material is present), and more complex heterostructures can be obtained by a sequence of injections. We discuss this approach first with $PEA_2PbBr_4$ as the core and an iodide phase as the frame by the delayed injection of $PEA_2PbI_4$ dissolved in GBL, and then show the extension to other materials and more complex structures in the SI in Section 2.4.

The hyperspectral confocal microscopy image in Figure 3b features a deep blue color of the core and a homogeneously green color of the frame. The transition from the green-emitting frame to the blue-emitting core is sharp, as evident from the extracted PL intensity profiles in **Fig. S10**, and single crystalline as demonstrated by the analysis of 4DSTEM diffraction patterns in **Figure S11**. Spatial 4D-STEM mapping (**Figure S11)** reveals that the interface region is restricted to less than 150 nm. Notably, the interface remains well-defined even after extended storage (with no detectable change in the PL peak position (**Fig. S12**)), and heating up to 100°C (**Figure S13**). The high color purity is confirmed by the PL spectra in Figure 3c that feature distinct PL peaks centered at 411 nm and 513 nm, thus at spectral positions that match the pure bromide phase and an iodide phase with a small Br content, respectively. Similarly, the UV-Vis absorption spectra show the features of phase-pure $PEA_2PbBr_4$, while the absorption edge of the iodide phase is slightly blue-shifted with respect to that of $PEA_2PbI_4$. The sequential injection approach also offers the opportunity to tune the composition of the frame material, and thus its band gap, namely by injecting mixed-halide solutions with varying $x_I$=[I]/([I]+[Br]) content (**Table S4**, **Fig. S14-15** and Figure 3c-d). The absorption edge and PL band of the frame shift progressively toward longer wavelengths with increasing $x_I$, demonstrating that the band gap of the frame material can be tuned via the halide ratio in the injected solution. Interestingly, regardless of iodide content in the frame, the spectral features of the $PEA_2PbBr_4$ core remain unchanged, confirming that this approach preserves the purity of the core material. SEM images of pristine and heterostructured microcrystals were recorded to evaluate the size distribution (**Fig. S16**). As expected, the heterostructures exhibit larger dimensions compared to the pristine Br-based crystals, and their size can be tuned by adjusting the antisolvent injection rate (in the range from few to several tens of micrometers). The small batch-to-batch variation of the size distribution further demonstrates the high reproducibility of this approach.

We also performed in situ PL measurements for the sequential injection strategy to monitor the crystallization kinetics (Figure 3e and **Fig. S17** for homo structures). Initially, we observe a weak band at 570 nm that we assign to $[PbBr_3]^-$-GBL complexes formed in solution (see **Fig. S18**). At 2 mL of antisolvent, the emission of crystalline $PEA_2PbBr_4$ emerges (at 410 nm), marking the onset of crystallization. The PL intensity sharply increases and then gradually saturates, consistent with the consumption of the precursors and completion of $PEA_2PbBr_4$ microcrystal formation. Subsequent injection of the $PEA_2PbI_4$ precursor solution (at 9 mL) in GBL results in an almost instantaneous drop in the $PEA_2PbBr_4$ PL intensity that we attribute to absorption of the $PEA_2PbBr_4$ emission by the injected $PEA_2PbI_4$ precursor (**Fig. S18**). At 12 mL of injected antisolvent, the PL band corresponding to the iodide-containing 2DLP phase appears at 500 nm, and gradually shifts to 510 nm. This indicates the growth of I:Br alloyed phases at the heterojunction with a decrease in Br content until a stable composition (with ~20% Br content obtained from EDX) is reached (at 20 mL) that emits at 510 nm. We note that with an optimized synthesis (where the iodide precursor was injected close to the crystallization point of $PEA_2PbI_4$ and combined with a high antisolvent injection rate) we reproducibly obtained heterostructures with 8% of Br content in the iodide frame, as discussed in detail in Section 3 in the SI (**Figures S19-S22**). The residual Br content in the frame could result from not complete consumption of the Br precursors in the prior core growth, and from interdiffusion of anions at the interface while a relatively I-pure-phase starts to crystallize on the edge of the $PEA_2PbBr_4$ cores.

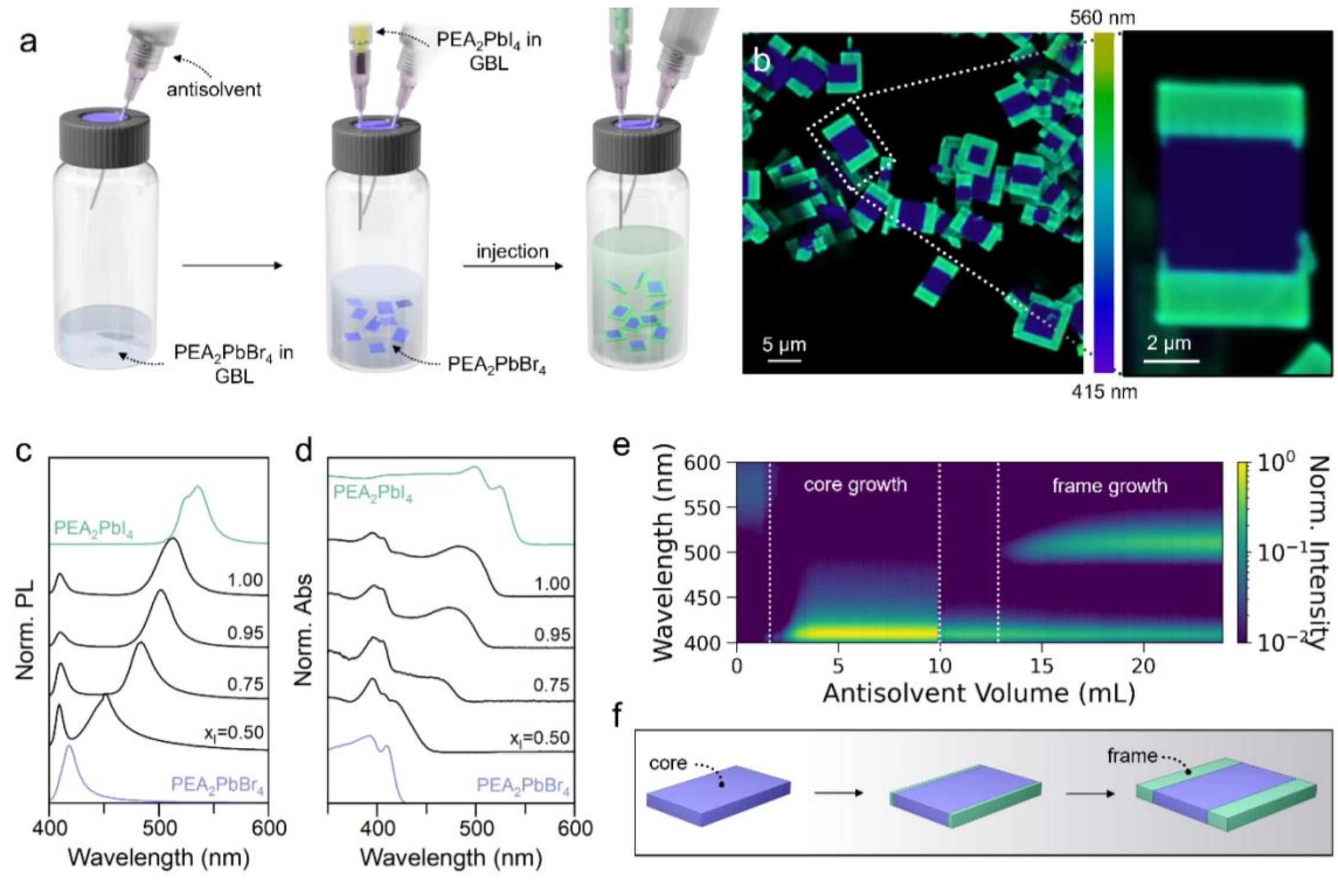

**Figure 3.** Sequential growth of heterostructures with sharp interfaces. a) Illustration of the sequential injection growth process. b) Confocal hyperspectral microscopy images showing $PEA_2PbBr_4/PEA_2PbBr_xI_{4-x}$ heterostructures in core-frame geometry that consist of pure Br phase in the core and slightly alloyed Br:I composition at the edges ($\lambda_{ex}$=400 nm). c-d) Absorption and PL spectra recorded from the heterostructure microcrystals grown by sequential injection with different frame compositions ($x_I$=[I]/([I]+[Br]) is the nominal composition). The spectra of pure phases ($PEA_2PbBr_4$ and $PEA_2PbI_4$ homostructures) are reported for comparison ($\lambda_{ex}$=370 nm). e) In-situ recording of PL emission ($\lambda_{ex}$=365 nm) during the sequential growth,

where the evolution is shown versus injected antisolvent volume (DCB, 10 mL $min^{-1}$). f) Scheme illustrating the growth process of the triptych-shaped heterostructures.

One key question is whether there is optical coupling between the core and frame regions, for example, if optical excitation of the higher band gap core region (with UV or blue light) can result in efficient green emission of the frame. Under wide-field UV light excitation (Figure 4a), the heterostructures exhibit blue and bright green emission from the core and the frame, respectively (Figure 4c and **Fig. S23**). We then illuminated the core of the heterostructures selectively (Figure 4b) with a pulsed femtosecond laser at 375 nm via a 100x objective (NA 1.3 oil) and recorded spatially and spectrally resolved images of the light emission, as depicted in Figure 4d and e. We clearly observe bright and homogeneous green emission from the frame regions, indicating efficient optical coupling of the core and frame. Selective excitation of one of the two iodide extremities results in no detectable emission from the bromide phase, confirming the directional nature of the coupling (**Fig. S24**). To gain deeper insight into the underlying mechanism, we performed time-resolved photoluminescence (TRPL) measurements on ensembles of heterostructures (see **Fig. S25 and Table S5-6**) and measured their photoluminescence quantum yield (PLQY), see **Fig. S26 and Table S7** for details. We find that the core-related blue emission decay is faster in the heterostructures with respect to the $PEA_2PbBr_4$ homostructures, and that also the blue emission intensity (PLQY) is reduced (from 14% to 1%), while the overall PLQY (blue and green) of the heterostructures is around 6%. This points to a loss channel with fast decay for the bromide phase recombination, but the strong emission intensity from the frame region indicates also efficient core to frame coupling. Indeed, for the green emission of the frames, we observe a longer lifetime in the heterostructures (compared to the pure microcrystals), and the green emission intensity is significantly increased (from 1% of the pure $PEA_2PbI_4$ crystals to 6% in the heterostructures). Therefore, the iodide phase radiative recombination is strongly increased, which we attribute to reabsorption and photon recycling of the core emission, supported by efficient waveguiding in microcrystals. which results in increased emission (higher PLQY compared to the homostructures) from the frame regions. This interpretation is supported by the reduced life time of the $PEA_2PbBr_4$ core compared to the homostructure **(Figure S 25a)**, since the long-lived radiative recombination (involved in the reabsorption) does not contribute anymore to the detection channel in the TRPL measurement. Furthermore, the longer lifetime of the iodide emission compared to that of the $PEA_2PbI_4$ homostructure microcrystals is in line with this reasoning. For more details on the optical coupling see **Figure S27** in Section 2.2 in the SI.

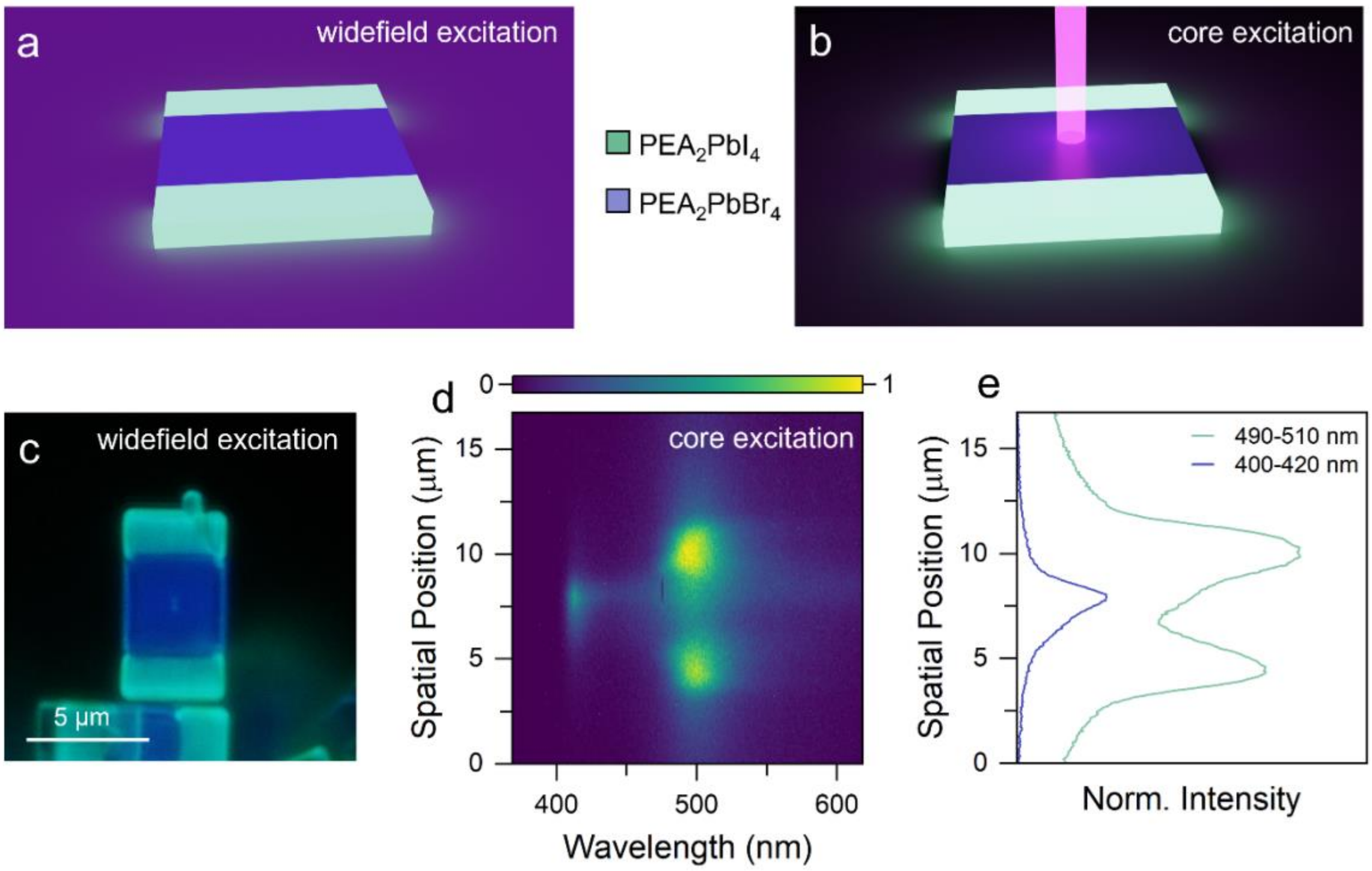

**Figure 4.** Optical coupling between the $PEA_2PbBr_4$ and $PEA_2PbI_4$ phases. (a, b) Schematic illustration of wide-field illumination and selective core excitation. (c) Optical emission image of a triptych-shaped heterostructure ($\lambda_{ex}$=365 nm). (d) Spatially resolved emission spectra of a similar heterostructure under selective excitation of the Br-based core ($\lambda_{ex}$=375 nm), showing emission from the $PEA_2PbI_4$ extremities. (e) Spatial intensity distributions of the $PEA_2PbBr_4$ emission (400–420 nm) and the $PEA_2PbI_4$ emission (490-510 nm) extracted from d).

The heterojunction interface obtained with the sequential injection is well-defined, as demonstrated by STEM-EDX elemental maps and the linescans in Figure 5a-e that show a sharp change in intensity for the Br/I signal and no iodine content in the core (see also Figure S11 for the change in lattice parameter at the interface). Electron diffraction patterns recorded with nanoscale resolution, obtained by integrating individual 4DSTEM pixels over homogeneous areas (Figure 5f-g) from regions spanning core to frame confirm continuous crystallinity (**Figure S11)** and are consistent with a monoclinic unit cell (**Table S1**). Interestingly, we observe two distinct populations of heterostructures, one with a core-frame morphology, and one with a triptych shape where the iodide phase grows only on two opposite edges of the core. The origin of these two different growth behaviors resides in the orientation of the octahedra lattice in the core, specifically if the edge terminations run along the more robust [100]/[010] or the 45°-rotated [$\bar{1}$10]/[110] directions, as illustrated in **Fig. S28**. The lattice orientation in the core also determines the edge termination, which is either in zig-zag ([100]/[010]) or armchair ([$\bar{1}$10]/[110]) configuration.[27] The appearance of microcrystals with corners truncated at 45° in the optical microscope images (**Fig. S28**) already evidences the possible occurrence of both edge terminations. 4D-STEM corroborates the two different lattice orientations (rotated by 45°) in the cores for these two populations (Figure 5f-g). In particular, the asymmetry responsible for the more frequently observed triptych-shaped structures under our standard growth conditions arises from the existence of two configurations of the octahedra within the inorganic layers. This distinction is linked to the possibility for the equatorial Br atoms, along [010] direction, to occupy two distinct crystallographic positions, as determined by XRD refinements (**Fig. S2 and Table S1**). In contrast, the apical Br atoms, which grow along [100] direction, were found to occupy a single position. Such a double octahedral configuration cannot occur when the lattice adopts a 45°-rotated orientation along the crystal edges - that is, when the crystal terminates along the [$\bar{1}$10]/[110] directions - because this orientation enforces a symmetric distribution of equatorial sites at the edges (Figure 5i). However, we found that the core-frame population can dominate when a fast frame growth is favoured in the fabrication process (see **Figures S29 and S30** in the SI).

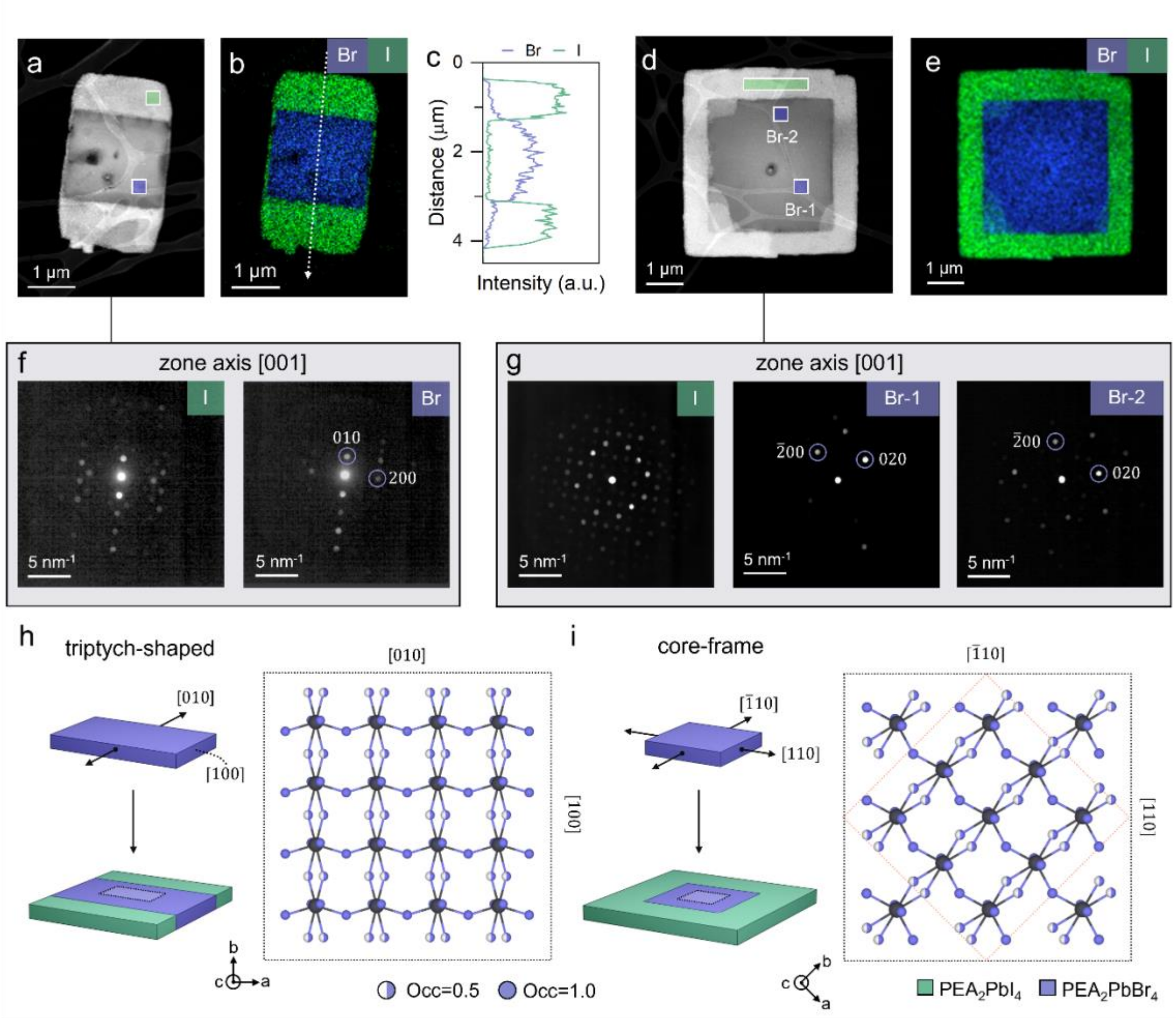

**Figure 5. Crystallographic origin of core–frame and triptych-shaped heterostructures.** (a) STEM image and (b) corresponding EDX map of a typical triptych-shaped heterostructure. (c) EDX line scan along the arrow in (b) showing the sharp change in Br and I signal at the interface. (d, e) STEM image and (e) corresponding EDX map of a core-frame heterostructure. (f) Integrated 4D-STEM patterns from the I side panels and Br core of the triptych structure marked in (a). (g) Integrated 4D-STEM patterns from the I frame and Br core of the core-frame structure marked in (d). The prominent reflexes in the electron diffraction patterns in f-g are highlighted by labels specifying the crystallographic direction. h) Illustration of the directional growth leading to the triptych structure and top view of the octahedra lattice of the core with [100]/[010] lattice direction. i) Scheme of the symmetric core-frame growth originating from a core with 45° rotated lattice.

Another highly appealing feature of the sequential injection method is that it enables the fabrication of multiple sequences of heterojunctions, such as triple halide microcrystals with decreasing band gap towards the edges that can lead to funneling of the photoexcited charges. We demonstrate such triple-halide architecture in Figure 1e and **Figures S31-33**, consisting of a Cl phase as the core, and a sequence of bromide phase and iodide phase frame and panels, respectively. Our one-pot synthesis method to obtain microcrystal heterostructures can furthermore be extended to "inverted structures" where the lower band gap material forms the core (**Figure S34 and S35**), and systems with different metal cations. We show the latter with $PEA_2PbBr_4$ core / $PEA_4AgBiBr_8$ frame heterostructures as depicted in Figure 1g and **Figure S36**.[35] We also managed to obtain "double frame" architectures that consist of an inner frame with higher Br and outer frame with lower content (**Figure S29**). Such multiple sequence injection can then be employed to grow an additional $PEA_2PbI_4$ frame, such that the thickness of the intermediate $PEA_4AgBiBr_8$ layer would control the optical coupling between the $PEA_2PbBr_4$ core and the outer $PEA_2PbI_4$ frame. Furthermore, we demonstrate in **Figure S37** that our heterostructure growth works also with other organic cations such as TEA (2-Thiophenethylamine) and BA (Butylamine).

### 2.3 Electric Potential Mapping with Kelvin Probe Force Microscopy

The main advantage of the capability to tailor the properties of lateral heterojunctions is the ability to design the potential landscape and to control charge and energy flow within the single microcrystals. We investigated the electrostatic potential of the heterostructure microcrystals using Kelvin Probe Force Microscopy (KPFM), which enables the evaluation of both the absolute work function and relative variations in electrostatic potential across interfaces (Figure S38).[45] The work function is closely linked to surface chemistry, doping, band bending and charge trapping [46], and therefore KPFM provides key information about the surface's chemical and electronic properties on the nanoscale. Here we focus on heterostructures with different halides (Cl, Br, I). In a first step, we measured a set of phase-pure 2D microcrystals for each halide and obtained the work function as listed in Table S9 and illustrated in Figure 6a (for details see Section 5 in the SI).

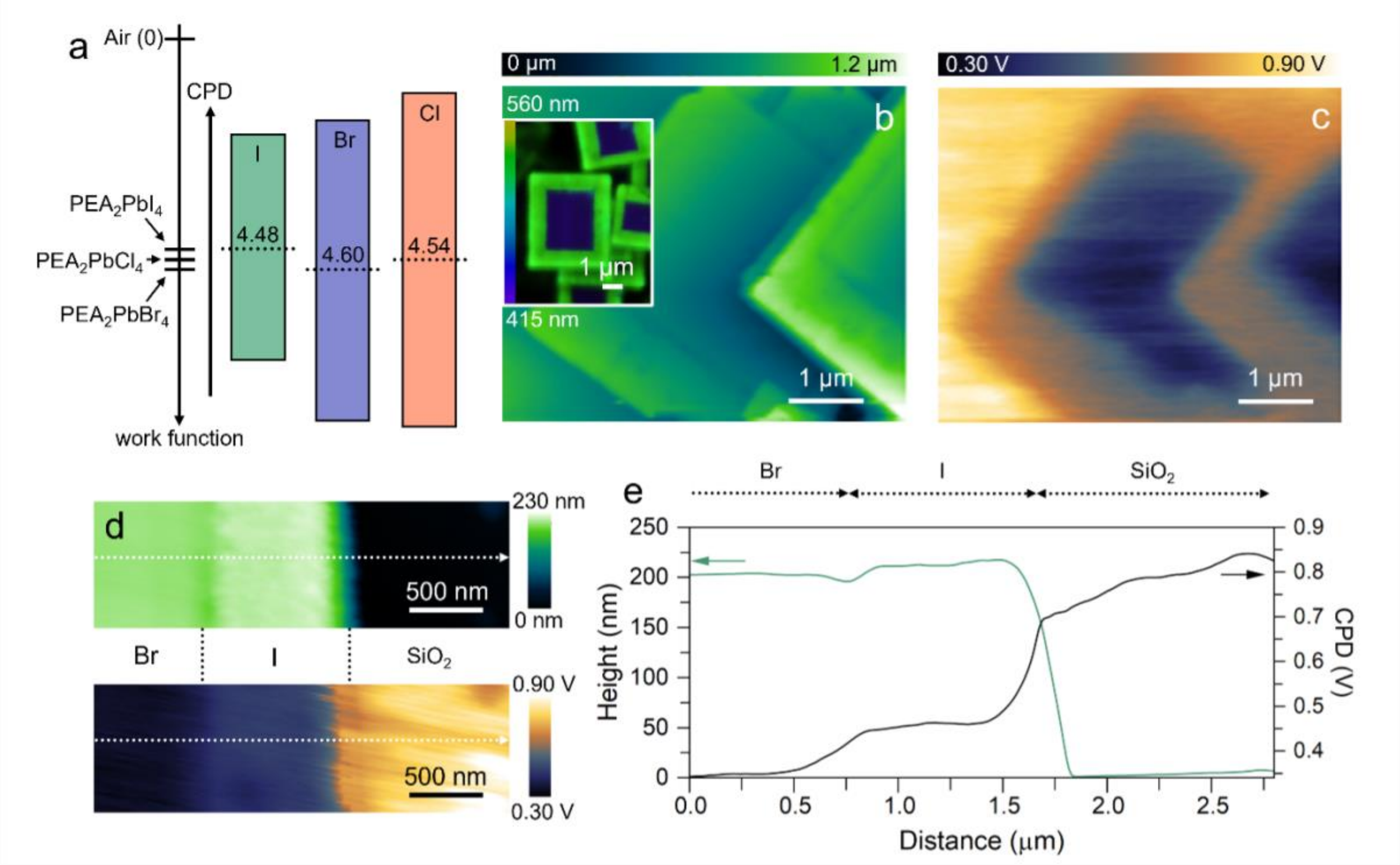

**Figure 6. Electrical potential landscape in microcrystal heterostructures fabricated by sequential injection.** a) Work functions of the 2D perovskites with pure halide phases measured by KPFM from homostructures, and illustration of the band alignment of the three different materials (assuming intrinsic semiconducting behavior) indicating the hypothesized charge transfer upon contact formation. The opposite polarity of Charge Potential Difference (CPD) with respect to the work function is illustrated by the arrows. The band gap taken from the excitonic peak in absorption is indicated by the height of the colored rectangles. b) AFM topography (inset shows a typical hyperspectral confocal fluorescence image of single

microcrystal from this batch), and c), CPD signal measured in KPFM of single core-frame heterostructure microcrystals. d) Topography and CPD measurements on a magnified region extending from the center to the edge of the microcrystals. e) Averaged height and CPD profiles obtained from the data in d). For the KPFM experiments, the microcrystals were deposited on $Si/SiO_2$ (150 nm) substrates.

We note that the order of increasing work function does not correlate with the increase in band gap of these materials. Figure 6b shows core-frame bromide-iodide microcrystal heterostructures fabricated by sequential injection, thus with $PEA_2PbBr_4$ core, and a nearly phase pure $PEA_2PbI_{4-x}$ frame that contains only a 7-8 % of bromide. The hyperspectral confocal PL images (inset of Fig. 6b) demonstrate the high quality of the heterostructures in terms of interface and phase homogeneity. Atomic force microscopy (AFM) and KPFM images of core-frame heterostructure microcrystals are reported in panels (b) and (c). The higher work function of the bromide phase (lower CPD signal) is evident in the core, and the frame manifest the higher CPD signal expected for the iodide phase. Figure 6d shows the iodide-bromide interface section on a magnified scale, revealing a small dip in height at the interface. The CPD profile shows a gradual increase from the bromide core to the iodide region that spans approximately 300-400 nm (Figure 6e), indicative of a charge redistribution that extends beyond the morphological interface. This finding highlights the complementarity of KPFM measurements with respect to structural investigations, revealing that the electrostatic interface region can span larger distances than the crystalline one. The relative decrease in CPD signal in the iodide region towards the interface could be rationalized by positive charge migration from the bromide to the iodide phases. Correspondingly, the lack of positive charges in the bromide at the interface region would explain the increase in CPD towards the interface. The band alignment illustrated in Figure 6a, with a larger offset in the valence band due to the ca 120 meV work function difference, is in agreement with such a charge migration process. Interestingly, we do not observe any indication of charge accumulation (that is localised valleys or peaks in CPD) at the interface that could be related to defects, vacancies or other charge traps. KPFM measurements on triptych-shaped $PEA_2PbBr_4$-$PEA_2PbI_{4-x}$ heterostructures are shown in **Figure S39**.

## 3. Conclusion

We introduce a novel method for the growth of lateral heterostructures in 2D metal-halide perovskite microcrystals via antisolvent triggered recrystallization that provides unprecedented design flexibility in terms of architecture and composition. Compositional control via the concentration of the dissolved materials in the growth solution allows to tune the band gap and the related offset at the interface, enabling pathways to tailor the emission colors, charge transfer, and recombination dynamics at the junction. Sequential injection of different materials during the reaction leads to nearly phase-pure materials, sharp and crystalline interfaces, and allows for freedom to tailor the material sequence of the heterojunctions, as well as the possibility to grow a series of multiple junctions. Our approach provides a versatile platform to design the energy band gap and the electrical potential landscape in perovskite materials that are highly appealing for optoelectronic applications such as photocatalysis, energy harvesting, and devices like LEDs and photodetectors. Furthermore, the basic concept of heterostructure fabrication by sequential recrystallization should be extendable to other material systems. For example, metal-organic chalcogenolates were recently shown to dissolve in polydentate amines, suggesting that similar solubility-driven strategies could enable their controlled heterostructure formation.[47]

Towards device applications, the ability to tune band offsets and guide energy flow within a single microcrystal opens pathways for luminescent solar concentrators and scintillator detectors that rely on materials in which emitted photons experience minimal reabsorption.[48–51] The efficient optical coupling to the lowest band gap region in our heterostructures ensure that emission occurs at energies that are not reabsorbed by higher-bandgap domains. Such suppression of self-absorption highlights the potential for developing next-generation light-management and radiation-detection materials.

## 4. Experimental Section

*Chemicals and Reagents:* Lead(II)chloride ($PbCl_2$, 98%), lead(II)bromide ($PbBr_2$, ≥98%), lead(II)iodide ($PbI_2$, 99%), silver bromide (AgBr, ≥99.99% trace metals basis), bismuth(III)bromide ($BiBr_3$, ≥98%), hydrochloric acid (HCl, 37%), hydrobromic acid (HBr, 48%), hydroiodic acid (HI, 57%, distilled, 99.999% trace metal basis), hypophosphorous acid ($H_3PO_2$, 50%), phenethylamine (PEA, ≥99%), Phenethylammonium bromide (PEABr, ≥98%), acetone (≥99.5%), ethyl acetate (≥99.5%), toluene (≥99.7%, anhydrous), γ-Butyrolactone (GBL, ≥99%), acetonitrile (ACN, anhydrous, 99.8 %), 1,2-dichlorobenzene (DCB, anhydrous, 99.8 %), diethyl ether (DEE, ≥99.7%), octane (anhydrous, ≥99%) were purchased from Sigma-Aldrich without any further purification.

*Synthesis of $PEA_2PbX_4$ Microcrystalline Powders*: $PEA_2PbBr_4$ microcrystalline powder was obtained using a previously published protocol with modifications.[43] Specifically, 92 mg (0.25 mmol) of $PbBr_2$ was dissolved in 200 μL of HBr, followed by dilution with 2 mL of acetone. Subsequently, 75 μL (0.6 mmol) of PEA was injected into the solution, resulting in the immediate formation of micrometer-sized $PEA_2PbBr_4$ microcrystals. The reaction mixture was continuously agitated for a minimum of 3 h to ensure a complete reaction. Afterward, the microcrystalline powder was separated from the solution by centrifugation at 4000 rpm for 2 min, followed by redispersion in 2 mL of acetone. The washing procedure was repeated at least two more times, and the purified powders were dried under vacuum for 2 h. For the synthesis of $PEA_2PbI_4$ microcrystalline powder, the same procedure was applied, with the substitution of $PbBr_2$ and HBr for $PbI_2$ (115 mg, 0.25 mmol) and a mixture of HI (200 μL) and $H_3PO_2$ (108 μL), respectively. Furthermore, to account for the higher solubility of $PEA_2PbI_4$ in acetone, ethyl acetate was used as antisolvent. $PEA_2PbCl_4$ microcrystalline powder where prepared, dissolving 69.52 mg (0.25 mmol) of $PbCl_2$ in 200 μL of HCl, followed by the dilution with 2 mL of acetone. Then, 75 μL (0.6 mmol) of PEA was added to the solution, forming $PEA_2PbCl_4$ microcrystalline powders. The mixture was continuously agitated for at least 12 h. The subsequent washing and drying process is the same as for $PEA_2PbBr_4$.

*Synthesis of $PEA_4AgBiBr_8$ Microcrystalline Powders*: $PEA_2AgBiBr_8$ powders were prepared based on a previously published method with minor modifications.[52] Typically, 0.5 mmol of AgBr, 0.5 mmol of $BiBr_3$, and 2 mmol of PEABr were dissolved in 5 mL of concentrated HBr at 120 °C under continuous stirring. After the mixture turned clear, the heating was stopped, and the solution was allowed to cool naturally to room temperature, during which large yellow crystals formed. The crystals were isolated by vacuum filtration, rinsed with cold diethyl ether, and dried under vacuum for 2 h.

*Preparation of 2DLP stock solutions:* Stock solutions of the 2DLPs were prepared by dissolving a certain amount of the corresponding powder in the desired solvent (see Table S2). The mixtures were stirred for at least 2 h at room temperature, then filtered through a 450 nm PTFE syringe filter and stored for further use.

*General description of the preparation of homo and heterostructures:* The general procedure for preparing homo- and heterostructures is as follows. 2DLP stock solutions were transferred into a 40 mL vial equipped with a stir bar. For homostructures, a defined volume of a single 2DLP stock solution was added to the vial prior to antisolvent injection. For heterostructures with a gradient alloyed junction, the $PEA_2PbBr_4$ and $PEA_2PbI_4$ precursor solutions were mixed in the desired ratio prior to antisolvent injection. For heterostructures with a phase-pure core, the precursor solution of the core material was placed in the vial first, followed by antisolvent injection. A defined volume of the stock solution of the second phase was added only after the desired antisolvent volume had been introduced under constant stirring. In all cases, up to 24 mL of antisolvent were injected at a constant rate (typically 1 mL $min^{-1}$ or 10 mL $min^{-1}$). After the antisolvent addition was complete, the mixture was stirred for an additional 5 min, then centrifuged at 4000 rpm for 2 min. The supernatant was discarded, and 10 mL of the respective antisolvent were added to redisperse the precipitate. This washing step was repeated two more times. Finally, the homo- or heterostructures were dispersed in octane inside a glovebox and stored for further use.

*Gradient alloyed heterostructures from mixed precursor solutions:* 3 mL of a $PEA_2PbBr_4$ stock solution in ACN were mixed with a defined amount of a $PEA_2PbI_4$ stock solution in ACN to achieve the desired ratio (see Table S3). Up to 24 mL of antisolvent (toluene or DCB) were then added at 1 mL $min^{-1}$ to trigger crystallization.

*$PEA_2PbBr_4$-$PEA_2PbI_4$ double-halide heterostructure:* 1.2 mL of a $PEA_2PbBr_4$ stock solution in GBL was added to a 40 mL vial, and DCB was injected as the antisolvent under constant stirring. After 10 mL of antisolvent had been introduced, 0.15 mL of the $PEA_2PbI_4$-GBL stock solution was swiftly injected. The antisolvent injection continued until the full 24 mL of DCB had been added. To tune the composition of the second phase, $PEA_2PbBr_4$ and $PEA_2PbI_4$ stock solutions were premixed in the desired ratio prior to injection (Table S4).

*$PEA_2PbCl_4$-$PEA_2PbBr_4$ double-halide heterostructure:* A total of 0.6 mL of the $PEA_2PbCl_4$–GBL stock solution was added to a 40 mL vial, and DCB was injected as the antisolvent using a syringe pump at 1 mL $min^{-1}$. After 5 minutes, once $PEA_2PbCl_4$ core had formed, 1.2 mL of the $PEA_2PbBr_4$–GBL stock solution was swiftly injected, and the antisolvent addition continued at the same rate until the full 24 mL of DCB had been injected.

*Synthesis of triple halide heterostructures of the form $PEA_2PbCl_4$-$PEA_2PbBr_4$-$PEA_2PbI_4$:* To initiate heterostructure formation, 0.6 mL of the $PEA_2PbCl_4$ stock solution was added to a 40 mL vial. Under constant stirring, DCB was introduced as the antisolvent at a rate of 1 mL $min^{-1}$. After 5 mL of DCB had been added, 1.2 mL of the $PEA_2PbBr_4$ stock solution was injected, forming the first frame of the heterostructure. For the third phase, 0.15 mL of the $PEA_2PbI_4$ stock solution was added after a total of 15 mL of DCB had been introduced. Antisolvent injection continued until the full 24 mL was injected.

*$PEA_2PbBr_4$-$PEA_4AgBiBr_8$ heterostructure:* 3 mL of the $PEA_2PbBr_4$ stock solution in ACN were added to a 40 mL vial, and toluene was introduced as the antisolvent at a rate of 1 mL $min^{-1}$. After 5 mL of toluene had been injected and the $PEA_2PbBr_4$ core had formed, 1 mL of the $PEA_4AgBiBr_8$ stock solution in ACN was swiftly added. The toluene injection then continued until a total of 24 mL had been delivered.

*Solubility of 2DLPs in ACN:* To determine the solubility, 50 mg of the 2DLP powder was dispersed in 4 mL of ACN and stirred for 48 h at 25 °C. The mixture was then centrifuged at 4000 rpm for 2 min, and the supernatant was filtered through a 450 nm PTFE syringe filter. A 50 µL aliquot of the filtered solution was transferred to a Falcon tube and allowed to evaporate to dryness. After evaporation, 1 mL of aqua regia ($HCl/HNO_3$=3/1, v/v) was added to dissolve the recrystallized material overnight, followed by dilution with 9 mL of Milli-Q water. The resulting solution was filtered again using a 450 nm regenerated cellulose syringe filter. Elemental quantification was performed by inductively coupled plasma optical emission spectroscopy (ICP-OES) using an iCAP 6300 DUO spectrometer.

*Optical Characterization:* The PL spectra were acquired with an Edinburgh Instruments (FLS920) fluorescence spectrometer equipped with a Xenon lamp and monochromator for steady-state PL excitation.PLQY values of the samples were measured in an integrating sphere attached to the same instrument with an excitation wavelength of 370 nm or 450 nm. Time-resolved PL measurements were carried out with a time-correlated single-photon counting (TCSPC) unit coupled to a pulsed diode laser. The samples were excited at 372 nm with 50 ps pulses at a repetition rate of 1 µs. The absorption spectra were collected from the samples using a Varian Cary 5000 ultraviolet–visible–near-infrared (UV-Vis-NIR) spectrophotometer equipped with an external diffuse reflectance accessory, and operating in absorption geometry. The hyperspectral confocal images were collected with a Nikon A1R+/A1+ confocal laser microscope equipped with a 60x oil immersion objective. The samples were excited with a 400 nm laser. The PL was collected in the range of 410 nm and 560 nm with a resolution of 6 nm. For all measurements described above, the samples dispersed in octane were typically drop-cast onto thin glass coverslips (0.17 mm) or silica substrates.

*In situ PL measurements:* The measurements were performed by directing a 365 nm LED (Thorlabs) in free space into a 40 mL vial containing the 2DLP precursor solution under constant stirring. The spot was slightly defocused to excite a larger sample volume to reduce the effect of intensity fluctuation due to the scattering of the formed microcrystals. The PL signal was collected by an 20 mm lens at a 90° angle using a fiber-coupled spectrometer (AvaSpec2048, Avantes). When the initial solution volume was insufficient for homogeneous free-space excitation, light from a 365 nm fiber-coupled LED (Thorlabs) was delivered through a bifurcated fiber inserted into a glass test tube. The test tube containing the fiber was then immersed directly into a 50 mL flask containing the precursor solution. Before entering the spectrometer, the collected light was filtered using a 400 nm long-pass filter to prevent excitation light from reaching the detector. Depending on the antisolvent injection rate, the integration time of the spectrometer was set to 2 s (1 mL $min^{-1}$) or 0.2 s (10 mL $min^{-1}$). Before initiating the antisolvent injection, dark spectra were collected and subsequently used to correct all PL spectra acquired during the crystallization process. To suppress UV-light-induced oxidation of $I^-$ to $I_2/I_3^-$, 300 µL of dodecanethiol (DDT) was added to the antisolvent. The DDT effectively reduced $I_2/I_3^-$ species while forming the corresponding dithiol.

*Local selective excitation:* The optical excitation was provided with an 80 MHz pulsed Chameleon OPO laser with tunable wavelength (SHG: 340-540 nm). The excitation light (375 nm) was coupled to a custom-built µ-PL setup and focused on the sample with a 100x (NA 1.3) objective. The PL was imaged using Andor-CMOS camera. For imaging, the PL light was focused on the slit using a 200 mm lens. For the measurements the heterostructures dispersed in octane were drop-cast onto $Si/SiO_2$ (150 nm) substrates for the measurements.

*X-ray diffraction:* XRD patterns were collected on a PANalytical Empyrean X-ray diffractometer equipped with a 1.8 kW CuKα ceramic X-ray tube and a PIXcel3D 2×2 area detector, operating at 45 kV and 40 mA. The diffraction patterns were recorded under ambient conditions using a parallel beam geometry and in symmetric reflection mode. The homo- and heterostructures obtained from sequential crystallization, dispersed in octanol, were drop-cast onto a zero-diffraction silicon substrate for measurement.The XRD patterns of microcrystalline powder were recorded on a Malvern-PANalytical 3rd generation Empyrean X-ray diffractometer, equipped with a 2.5 kW MoKα X-ray tube (operating at 60 kV and 40 mA) and 1.8 kW CuKα X-ray tube (operating at 40 kV and 45 mA). The microcrystalline powder samples were measured in Debye-Scherrer (transmission) configuration, using a borosilicate glass capillary of 0.5 mm. The XRD pattern with Cu-source was collected from 4° to 90° with a step size of 0.014°, while the pattern with Mo-source was from 1.26° to 60° with a step size of 0.014°, using a focusing mirror, a capillary spinner sample stage and GaliPIX3D solid-state pixel detector acquiring in 1D mode.

*Atomic Force Microscopy:* AFM images and topological analysis of the heterostructures were performed using an XE-100 AFM system (Park Scientific, Suwon, South Korea), operating in a non-contact mode.

*Kelvin Probe Force Microscopy*: KPFM was conducted on an Asylum Research MFP-3D system inside a nitrogen glovebox (RH < 20%) to ensure environmental stability. ElectriMulti75E cantilevers (Budget Sensors; 70–75 kHz, ~3 N/m, 25 nm tip radius, 10–15 μm height) coated with PtIr ($\Phi \approx 5.02$ eV) were used. Topography was acquired in amplitude-modulation mode (20–40 nm oscillation). KPFM operated in dual-pass "NAP" mode: the first pass mapped topography, while the second measured the contact potential difference (VCPD) by applying an AC bias and nullifying the electrostatic force via a compensating DC voltage (VDC). Assuming a parallel-plate capacitor model:

$$V_{CPD} = (\Phi_s - \Phi_p)/e,$$

where $\Phi_s$ and $\Phi_p$ are the sample and probe work functions. Tip work functions were calibrated on freshly cleaved HOPG before measurements. AFM data were analyzed with Gwyddion. Homo- and heterostructures dispersed in octane were drop-cast onto $Si/SiO_2$ (150 nm) substrates for the measurements.

*Scanning Electron Microscopy and Energy Dispersive X-ray Spectroscopy:* Homo and heterostructures were investigated with a Zeiss GeminiSEM 560 (Zeiss, Oberkochen, Germany) scanning electron microscope equipped with a field-emission gun and operating at 10 kV acceleration voltage. The detector was a VPBSE, a back-scattered detector, used to enhance the differences of the elements present in the samples according to their brightness. The SEM is equipped with energy-dispersive spectroscopy (EDX, Oxford instrument, X-Max, 80 $mm^2$), and corresponding data were acquired at 30 kV. Samples were typically drop-casted on 10x10 mm $Si/SiO_2$ (150 nm) chips for investigation.

*Scanning Transmission Electron Microscopy:* Homo and heterostructures were drop cast onto Cu grids with a continuous or lacey carbon support film from an octane dispersion. High-Angle Annular Dark Field (HAADF) imaging and STEM- Energy-Dispersive X-ray (EDX) spectroscopy compositional analysis were carried out using an image-Cs-corrected JEOL JEM-2200FS with a Schottky emitter gun operated at 200 kV, equipped with a Bruker XFlash5060 silicon-drift EDX detector (SDD).

Further STEM images were acquired using a HAADF detector on a probe- and image-aberration-corrected ThermoFisher Spectra 300 S/TEM with an X-FEG source, operated at 300 kV. Elemental distribution maps and line scan profiles were recorded by collecting EDX signals on a Dual-X system, which comprises two EDX detectors positioned on either side of the sample.4D-STEM datasets were acquired on the same Spectra 300 in microprobe mode, at 300 kV and with a semi-convergence angle of 0.5 mrad, using a current of a few pA and a total dose <10 $e^-/A^2$. The diffraction patterns were recorded on a post-filter Gatan Continuum camera. 4D-STEM data cubes were analyzed in Gatan DigitalMicrograph, and diffraction patterns for indexing were simulated using CrysTBox.[53]

**Supporting Information**

Supporting Information is available free of charge online and contains:
Section 1. Two-dimensional layered perovskite powders and recrystallized microcrystal homostructures.
Section 2. Design and control on the alloyed phases in microcrystal heterostructures.
Section 3. Sequential injections of dissolved microcrystalline powders leading to nearly phase-pure heterostructures: interface characterization and optical coupling.
Section 4. Sequential growth with multiple injection leading to triple halide heterostructures.
Section 5. Mapping the electrostatic potential landscape with Kelvin Probe Force Microscopy.

**Acknowledgements**

A. S. acknowledges the European Union's Horizon 2020 research and innovation programme under the Marie Skłodowska-Curie Funding Program (Project Together, grant agreement No.101067869). R.K acknowledges funding by the European Union under Project 101131111 – DELIGHT. The authors thank Ryo Mizuta Graphics and Dr. Joseph G. Manion for providing the optical components pack and standard laboratory vials asset that was used for a part of the presented schemes. The authors thank Sergio Marras (Materials Characterization Facility, IIT) for assistance in the XRD data acquisition.

ToC figure

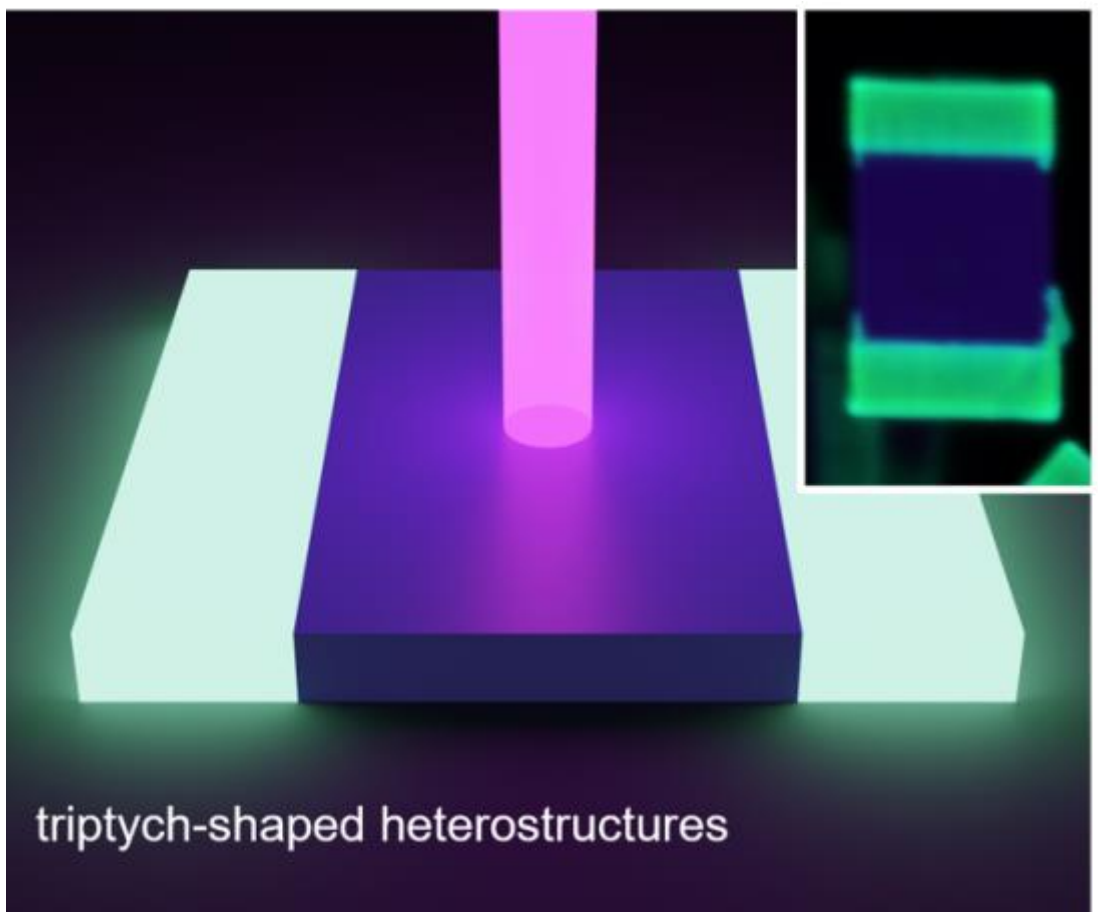